\documentstyle[rotate,epsf,psfig]{l-aa}
\thesaurus{ 06(02.16.2 ; 02.18.5 ; 02.18.7 ; 08.16.6 ; 08.16.7 PSR B0144+59, B0355+54)}

\begin{document}

\title{On the high frequency polarization of pulsar radio emission}

\author{ A. ~von Hoensbroech \inst{1},
J. ~Kijak \inst{2,1}
and A. ~Krawczyk \inst{2,1}}

\offprints {A. ~von~Hoensbroech (avh@mpifr-bonn.mpg.de)}

\institute{Max-Planck-Institut f\"ur Radioastronomie,
 Auf dem H\"ugel 69, D-53121 Bonn, Germany.
\and 
The Astronomical Centre of Zielona G\'ora, Lubuska 2, PL-65 265 Zielona G\'ora, Poland.}

\date{7 November 1997 ; 9 March 1998}

\maketitle
 
\markboth{~von Hoensbroech et al.: On the high frequency polarization of 
pulsar radio emission}{}

\begin{abstract}
We have analyzed the polarization properties of pulsars at 
an observing frequency of 4.9 GHz.
Together with low frequency data, we are able to trace polarization
profiles over more than three octaves into an interesting
frequency regime. At those high frequencies the polarization properties
often undergo important changes such as significant depolarization.
A detailed analysis allowed us to identify parameters, which
regulate those changes. A significant correlation was found between the
integrated degree of polarization and the loss of rotational energy $\dot E$.

The data were also used to review the widely established pulsar 
profile classification scheme of core- and cone-type beams.

We have discovered the existence of pulsars which show a strongly 
{\it increasing} degree of circular polarization towards high frequencies.

Previously unpublished average polarization profiles, recorded at the 
100m Effelsberg radio telescope, are presented for 32 radio pulsars
at 4.9 GHz. The data were used to derive polarimetric parameters and
emission heights.
\end{abstract}


\section{Introduction}

Polarimetry plays a key role in our understanding of the 
emission mechanism of pulsars, the ambient conditions in the
emission region and the geometrical structure of the magnetic
field. Similar to the great variety of profile shapes, 
the polarimetric features of the
radio emission vary strongly from pulsar to pulsar and from 
frequency to frequency. Virtually every polarization state
between totally unpolarized and fully linearly or highly 
circularly polarized can be found amongst different pulsars
and even within one profile. Also the shape of the polarization 
position angle (hereafter PPA) curve varies between nearly constant, 
a smooth orderly swing, sudden jumps and nearly chaotic behaviour.
The jumps in the PPA--swing often cover precisely $90^\circ$ and are
therefore called orthogonal polarization modes (hereafter OPM,
see e.g. Stinebring et al. 1984; Gil \& Lyne 1995; Gangadhara 1997).

Nevertheless certain common pulse features have been identified in the past, 
which lead to different classification attempts (e.g. Backer 1976; 
Rankin 1983; Lyne \& Manchester 1988).
It is widely accepted that two general types of profile components can
be identified: Those which are radiated from the outer parts 
of the emission tube as {\it conal} profile components and
those which are usually emitted from the central part as {\it core}-beams. 
This classification was initially
formulated systematically by Rankin (1983), for a detailed description we
refer to that paper, a short summary is given in Sect. \ref{types}. 
The identification of these components is mainly (but not only) based on the
frequency development of their polarization and their relative intensity.
In general this system has proven to be remarkably successful, although
in this paper we discuss some groups of pulsars which do not quite 
fit into this classification scheme.

One common polarimetric feature of most pulsars is the depolarization
towards high frequencies (in the following ``high frequency'' means
radio frequencies well above one GHz). Whereas the degree of polarization 
of pulsars is usually constantly high at low frequencies, it 
decreases rapidly above a certain frequency (e.g. Manchester 1971; 
Morris et al. 1981a; Xilouris et al. 1996). 
This tendency
is in contrast to the known properties of other astrophysical objects which 
have usually stronger polarization at higher frequencies, where the 
Faraday-depolarization effect is less severe.
Therefore, this effect is thought to be inherent to the pulsar magnetosphere, 
either intrinsic to the emission mechanism or due to a
propagation effect within the magnetosphere. The 
identification of the depolarization mechanism
is important as it might help to understand the environmental conditions
in the magnetosphere and the relevant physical processes. 
It is therefore necessary to carry out high frequency observations
as we would like to identify parameters which control this effect.

In this paper we also focus on the role of the circular polarization.
Theories proposed to explain this type of polarization
range from purely intrinsic mechanisms (e.g. \cite{RR90}) to pure propagation 
effects
(e.g. \cite{MS77}) and combinations of both (e.g. Kazbegi et al. 1991; 
Naik \& Kulkarni 1994). 
In order to distinguish between the different
mechanisms, it is necessary to trace the frequency development 
of the circular polarization over a large frequency interval.
Propagational effects should show a strong frequency dependence.

Whereas the degree of polarization varies strongly with frequency, 
the measured PPA 
is very stable over many octaves in frequency. If one allows for the 
occurrence of OPMs, the PPA swing is therefore thought to reflect 
the geometry of the pulsar magnetosphere as 
first noted by Radhakrishnan \& Cooke (1969). In some cases it is therefore 
possible to determine the viewing geometry of a pulsar by fitting the geometry 
dependent theoretical PPA curve -- the rotating vector model (hereafter RVM) 
-- to the measured PPA (the formula for the RVM is given e.g. 
by Manchester \& Taylor (1977)).

Many authors indicate the existence of a radius-to-frequency
mapping (hereafter RFM) where the radio emission is narrow band and
scales inversely with frequency (e.g. Cordes 1978; 
Blaskiewicz et al. 1991; Kramer et al. 1997;
Kijak \& Gil 1997; von Hoensbroech \& Xilouris 1997a). The knowledge of 
the existence and the strength of the RFM is important as it will
help to understand the emission physics. It is therefore necessary to
determine the emission height above the pulsar surface, where the
emission we observe at a certain frequency, originates. Polarimetry
provides one method amongst others to calculate this height.
This method was proposed by Blaskiewicz et al. (1991) and we have applied 
it to our data whenever possible (see Sect. \ref{Rem}).

Due to their steep radio spectra, pulsars tend to be rather weak sources 
at centimetre wavelengths. As a result, relatively little published 
data exists in this part of the spectrum
(Morris at al. 1981b; Xilouris et al. 1994; Xilouris et al. 1995; 
Manchester \& Johnston 1995; Xilouris et al. 1996; 
von Hoensbroech et al. 1997b).
In this paper we present the polarimetric properties of 32 weaker 
pulsars at 4.9 GHz which roughly doubles the number of published
polarization profiles at this frequency and allows statistical studies 
at such a high frequency for the first time.

\section{Observations}

\begin{table}
\caption[]{List of observations. The effective $\Delta t$ is the time resolution, 
which results from the temporal bin width and the dispersion smearing:
$\Delta t = \sqrt{\Delta t^2_{\rm bin}+\Delta t^2_{\rm disp}}$.}\label{table1}
\begin{tabular}{lllll}
\noalign{\smallskip}
\hline
\noalign{\smallskip}
PSR&	Date	&	Period&Pulses&eff. $\Delta t$\\
      &	[d-m-y]	&	[s]&No.	   &	[msec]\\
\noalign{\smallskip}
\hline
\noalign{\smallskip}
B0031$-$07&	16-08-95&	0.943&	600&	2.9\\
B0144+59&	08-11-95&	0.196&	4332&	1.4\\
B0402+61&	20-08-95&	0.595&	1000&	2.7\\
B0450$-$18&	21-08-95&	0.549&	540&	1.8\\
B0559$-$05&	08-11-95&	0.396&	5624&	3.2\\
\noalign{\smallskip}
B0611+22&	26-09-95&	0.335&	2244&	3.7\\
B0626+24&	28-09-96&	0.476&	3565&	3.4\\
B0628$-$28&	27-07-95&	1.244&	2349&	5.0\\
B0818$-$13&	20-08-95&	1.238&	480&	3.1\\
B0834+06&	26-09-95&	1.274&	1782&	2.8\\
\noalign{\smallskip}
B0906$-$17&	08-11-95&	0.402&	4255&	1.7\\
B0942$-$13&	08-11-95&	0.570&	5538&	1.3\\
B1039$-$19&	08-11-95&	1.386&	1180&	3.2\\
B1254$-$10&	20-08-95&	0.617&	960&	2.2\\
B1604$-$00&	15-08-95&	0.422&	700&	0.4\\
\noalign{\smallskip}
B1702$-$19&	26-08-95&	0.299&	1300&	0.8\\
B1706$-$16&	20-09-95&	0.653&	1833&	1.1\\
B1737+13&	20-08-95&	0.803&	738&	2.5\\
B1737$-$30&	02-09-91&	0.607&	960&	3.7\\
B1800$-$21&	16-08-95&	0.134&	3360&	8.5\\
\noalign{\smallskip}
B1821+05&	20-08-95&	0.753&	779&	3.9\\
B1823$-$13&	18-07-96&	0.101&	5880&	8.4\\
B1831+04&	14-08-95&	0.290&	2040&	3.2\\
B1839+56&	16-08-95&	1.653&	360&	3.7\\
B1911$-$04&	14-08-95&	0.826&	720&	3.4\\
\noalign{\smallskip}
B1913+10&	21-08-95&	0.405&	1702&	9.2\\
B1944+17&	20-09-95&	0.441&	1996&	2.7\\
B1952+29&	22-08-95&	0.427&	2800&	1.0\\
B2110+27&	21-08-95&	2.517&	984&	5.8\\
B2111+46&	09-11-95&	1.015&	2854&	5.6\\
\noalign{\smallskip}
B2303+30&	21-08-95&	1.576&	378&	3.9\\
B2334+61&	20-08-95&	0.495&	1200&	2.6\\
\noalign{\smallskip}
\hline
\noalign{\smallskip}
\end{tabular}
\end{table}

The observations were carried out at the Effelsberg 100-meter
radio telescope of the Max-Planck-Institute f\"ur Radioastronomie. 
The centre frequency was
set to 4.85 GHz with a bandwidth of 500 MHz and a system temperature 
of $T_{\rm sys}\simeq 30 $K. The left-hand (LHC) and
right-hand circular (RHC) output signals of the 
secondary-focus receiver were combined and detected in a broad-band 
multiplying polarimeter in the focus cabin. After an 
analogue-to-digital conversion using fast voltage-to-frequency converters, 
the signal was passed down to the 
pulsar backend where it was recorded synchronously to the topocentric
pulsar period. A detailed description 
of the Effelsberg pulsar observing system (EPOS) can be found in
\cite{EPOS}. The data were then dynamically calibrated using a calibration
signal which was injected synchronously to the pulsar period into the
waveguide. The full calibration method is described in
von Hoensbroech \& Xilouris (1997b).

The details of the observations are listed in Table \ref{table1}.

\section{Results}

\begin{table}
\caption[]{
Degrees of polarization, averaged over the whole pulse. 
The corresponding polarization diagrams are displayed in Figs. \ref{data1} to \ref{data4}.
$\bar \Pi_{\rm L}$ is the linear polarization, $| \bar \Pi_{\rm C}|$ 
and $\bar \Pi_{\rm C}$ represent the absolute value and the averaged
value of the circular polarization respectively (the convention is 
{\it positive} for
left-hand and {\it negative} for right-hand-sense. For B1702-19, 
values are given for the main pulse (MP) and the interpulse (IP).
}\label{table2}
\begin{tabular}{llll}
\noalign{\smallskip}
\hline
\noalign{\smallskip}
PSR&	$\bar \Pi_{\rm L}\pm\Delta_{\Pi_{\rm L}}$&$| \bar \Pi_{\rm C}|\pm\Delta_{ |\Pi_{\rm C}|}$&$\bar \Pi_{\rm C}\pm\Delta_{ \Pi_{\rm C}}$ \\
      &	[\%]	&[\%]	&[\%]\\
\noalign{\smallskip}
\hline
\noalign{\smallskip}
B0031$-$07&$11\pm1$	&$8\pm1$	&$+8\pm 1$	\\
B0144+59&$11\pm1$	&$50\pm1$	&$-49\pm1$	\\
B0402+61&$7\pm4$	&$16\pm4$	&$-2\pm6$	\\
B0450$-$18&$19\pm2$	&$8\pm2$	&$+7\pm4$	\\
B0559$-$05&$40\pm6$	&$10\pm3$	&$+5\pm5$	\\
\noalign{\smallskip}
B0611+22&$43\pm8$	&$27\pm6$	&$+26\pm9$	\\
B0626+24&$23\pm2$	&$9\pm2$	&$-9\pm3$	\\
B0628$-$28&$2\pm1$	&$3\pm1$	&$-3\pm2$	\\
B0818$-$13&$21\pm4$	&$12\pm4$	&$-3\pm6$	\\
B0834+06&$2\pm11$	&$23\pm8$	&$-5\pm12$	\\
\noalign{\smallskip}
B0906$-$17&$4\pm2$	&$6\pm2$	&$-2\pm3$	\\
B0942$-$13&$4\pm8$	&$8\pm7$	&$+1\pm11$	\\
B1039$-$19&$5\pm1$	&$8\pm2$	&$+6\pm2$	\\
B1254$-$10&$20\pm5$	&$12\pm4$	&$-8\pm7$	\\
B1604$-$00&$7\pm1$	&$7\pm1$	&$-4\pm1$	\\
\noalign{\smallskip}
B1702$-$19$^{\rm MP}$&$37\pm2$	&$14\pm2$	&$-10\pm3$	\\
B1702$-$19$^{\rm IP}$&$60\pm17$	&$17\pm13$	&$-6\pm21$	\\
B1706$-$16&$18\pm1$	&$5\pm1$	&$+0\pm1$	\\
B1737+13&$18\pm3$	&$9\pm2$	&$+4\pm4$	\\
B1737$-$30&$53\pm7$	&$59\pm4$	&$+59\pm5$	\\
\noalign{\smallskip}
B1800$-$21&$71\pm2$	&$33\pm1$	&$+33\pm2$	\\
B1821+05&$13\pm2$	&$10\pm2$	&$-1\pm3  $	\\
B1823$-$13&$70\pm3$	&$42\pm2$	&$+42\pm2$	\\
B1831+04&$15\pm1$	&$8\pm1$	&$-1\pm1$	\\
B1839+56&$10\pm2$	&$6\pm2$	&$+4\pm3$	\\
\noalign{\smallskip}
B1911$-$04&$8\pm2$	&$7\pm2$	&$-3\pm3$	\\
B1913+10&$4\pm7$	&$73\pm10$	&$+73\pm14$	\\
B1944+17&$18\pm2$	&$16\pm1$	&$-15.0\pm2$	\\
B1952+29&$8\pm1$	&$8\pm1$	&$+1\pm1$	\\
B2110+27&$22\pm6$	&$17\pm7$	&$+4\pm10$	\\
\noalign{\smallskip}
B2111+46&$22\pm1$	&$11\pm1$	&$+3\pm1$	\\
B2303+30&$17\pm2$	&$6\pm2$	&$+6\pm3$	\\
B2334+61&	$57\pm6$&	$18\pm4$&	$-8\pm7$\\
\noalign{\smallskip}
\hline
\noalign{\smallskip}
\end{tabular}
\end{table}

\begin{table}
\caption[]{Derived geometrical parameters of eight pulsars. $\alpha$ is the 
inclination angle between the rotation axis and the magnetic field axis, 
$\beta$ the {\it impact angle} between the magnetic field and the line of 
sight. $r_{\rm pol}$ is the emission height calculated using a polarimetric 
method (\cite{bcw}), $r_{\rm KG}$ calculated using a relationship proposed by 
Kijak \& Gil (1997), see Eq. \ref{KGeq}.}\label{table3}
\begin{tabular}{lllll}
\noalign{\smallskip}
\hline
\noalign{\smallskip}
PSR&	$\alpha [ ^\circ$]&$\beta [ ^\circ$]&$r_{\rm pol}$ [km]&$r_{\rm KG}$ [km]\\
\noalign{\smallskip}
\hline
\noalign{\smallskip}
B0144+59&$90\pm90$	&$0\pm90$	&$20\pm70$	&$190\pm90$\\
B1604$-$00&$90\pm75$	&$0.2\pm0.2$	&$380\pm80$	&$240\pm80$\\
B1702$-$19$^{\rm MP}$&$86\pm2$	&$-4\pm2$	&$300\pm90$	&$260\pm80$\\
B1702$-$19$^{\rm IP}$&	&		&$310\pm160$	&$260\pm80$\\
B1706$-$16&$90\pm65$	&$-2\pm1$	&$560\pm70$	&$330\pm100$\\
\noalign{\smallskip}
B1737+13&$90\pm86$	&$-2\pm2$	&$30\pm280$	&$320\pm110$\\
B1800$-$21&$90\pm90$	&$0\pm90$	&$-40\pm230$	&$190\pm50$\\
B2111+46&$38\pm37$	&$-4\pm4$	&$510\pm510$	&$320\pm150$\\
B2334+61&$90\pm88$	&$-8\pm7$	&$340\pm540$	&$390\pm50$\\
\noalign{\smallskip}
\hline
\noalign{\smallskip}
\end{tabular}
\end{table}

\vspace*{1cm}
\begin{figure}
\epsfysize5cm
\rotate[r]{\epsffile[160 70 510 420]{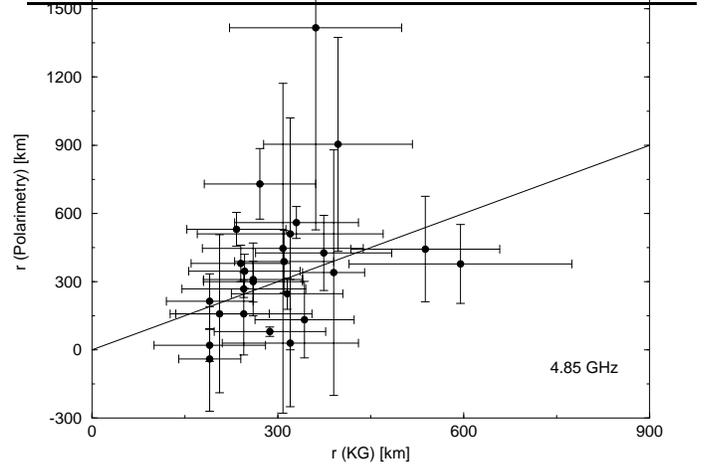}}
\caption{Comparison of emission heights derived with the polarimetry
method ($r_{\rm pol}$) and those derived with an
empirical relationship ($r_{\rm KG}$, see text). Those points whose 
error-bars cut the diagonal are consistent within their one $\sigma$
errors.}
\end{figure}

In this paper we present 32 average polarization profiles for weaker
radio pulsars as a result of a large survey at a wavelength of $\lambda$=
6.1 cm (\cite{K98}). 
The polarization diagrams are displayed in Figs. \ref{data1} to \ref{data4}.
In Table~2, we list the degrees of linear and circular
polarization for each source. Most of the pulsars in our sample show a fairly
low degree of polarization, which made the determination of the PPA
sometimes uncertain. 
Therefore the RVM could be fitted only for eight pulsars.
The fitted PPA swings are included into the Figs. \ref{data1} to \ref{data4}
for those eight pulsars.
The viewing geometry 
-- which is the inclination angle between the rotation axis and the 
magnetic field axis $\alpha$, and the ``impact angle'' between the magnetic 
field and the line of sight $\beta$ -- 
was determined for these objects.
The results for $\alpha$ and $\beta$
are listed in columns 2 and  3 of Table~3, respectively, with their 
estimated errors. These errors are sometimes relatively large due to the
ambiguity in the determination of the viewing geometry 
(see also \cite{avh1}). 
However, within our sample of 32 pulsars, there are some particularly 
interesting ones, which 
are discussed in the following sections.

\begin{figure*}[t]
\epsfysize5cm
\rotate[r]{\epsffile[300 25 560 235]{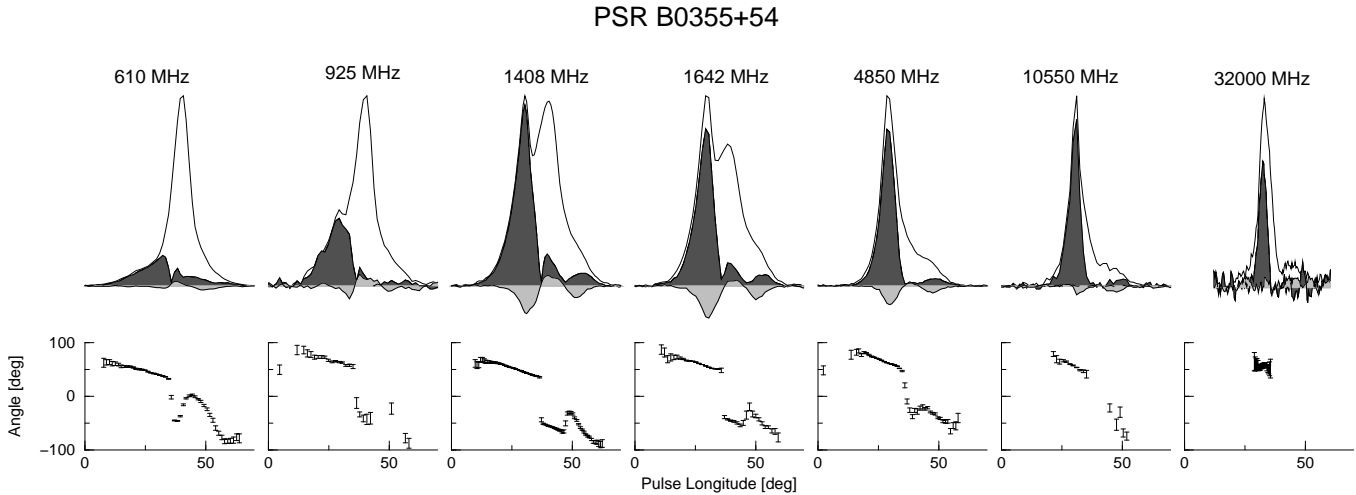}}
\caption[]{Polarization profiles of PSR B0355+54 
(profiles below 2 GHz are from Gould \& Lyne (1998), the others are 
Effelsberg data. The 32 GHz profile is from Xilouris et al. (1996)). 
The dark-shaded 
area represents the linear, the light-shaded area corresponds to the circularly
polarized intensity ({\it positive} = left-hand, 
{\it negative} = right-hand).
 Total power is represented by the unshaded solid line.
The first component is highly linearly polarized and 
becomes increasingly prominent at high frequencies.}\label{0355_freq}
\end{figure*}

\subsection{Interesting individual Objects}

\subsubsection{Interpulse: B1702-19}
At 4.9 GHz this pulsar is still highly polarized 
in both its main pulse and interpulse.
The main pulse appears to consist of  two or three 
components and is 50\% linearly
polarized. It shows right and left hand circular polarization.
The interpulse is totally linearly polarized but this linear polarization 
decreases to lower frequencies.  
Whereas the ratio of interpulse to main pulse 
intensity increases with frequency up to 1.4 GHz 
(see Biggs et al. 1988; Biggs 1993; Gould 1994), we find a decrease of this 
ratio above 1.4 GHz.
The geometrical parameters $\alpha$ and $\beta$ obtained from fitting PPA
resemble very much those obtained at lower frequencies 
using different methods (see Biggs et al. 1988; Lyne \& Manchester 1988).

\subsubsection{Wide profiles: B1800-21, B1823-13 and B1831-04}
With more than $100^\circ$ width in pulse longitude at this frequency, 
these pulsars have very wide profiles.
PSR B1800-21 is highly polarized and shows one sense of circular polarization.
The PPA follows an S-shaped course through the pulse.
PSR B1821-13 is similar to the former one and appears have two or more 
components in its profile. It is also highly polarized. 
A detailed discussion on these two pulsars will be given in Sect. \ref{1800}.
PSR B1831-04 has a multicomponent profile but low polarization. 

\subsubsection{High circular polarization: B0144+59, B1737-30 and B1913+10}
These objects show an unusually high degree of circular polarization 
($\geq 50\%$) at 4.9 GHz which is not seen at lower frequencies. 
A detailed discussion on the frequency 
development of these pulsars and the consequences of this observation 
is given in Sects. \ref{0144} and \ref{circ}.

\subsection{Geometry and emission altitudes}\label{Rem}

Radhakrishnan \& Cooke (1969) were the first to argue that the magnetic field
in the emission region has a dipolar structure and that the PPA corresponds
to the projection of the magnetic field line direction on the sky.
More recently, Blaskiewicz et al. (1991)
demonstrated that the PPA curves behave as predicted in the case
of a {\it vacuo} purely dipolar aligned rotator (\cite{GJ69}).
The relativistic model for the RVM (see \cite{bcw}) gives a formula for 
the calculation of absolute emission heights as
\begin{equation}
r_{\rm em}(\nu)= \frac{c}{4}\cdot \Delta t(\nu)~,
\end{equation}
where $\Delta t$ is the time lag between the centroid of the total intensity 
profile and the pulse phase at which the derivative of the PPA curve has its 
maximum. We calculated the emission altitudes for
nine profiles (eight pulsars, one with interpulse) at 4.9 GHz 
using our polarization data (see Table \ref{table3}). 
Together with the emission heights for 17 pulsars 
from (von Hoensbroech \& Xilouris 1997a) at the same frequency 
and derived using the same method, we compared the results with 
predictions from the relationship proposed by Kijak \& Gil (1997) 
\begin{equation}\label{KGeq}
r_{\rm KG}=(55\pm 5)\cdot R~\nu^{-0.21\pm 0.07}~\tau_{6}^{-0.07\pm 0.03}
P^{0.33\pm 0.05}\; .
\end{equation}
$R=10^4$m is the neutron star radius, $\nu$ is the frequency in GHz,
$\tau_6$ is the timing age in units of million years and $P$ is pulsar period 
in seconds. 
In most cases the results are consistent within the errors (see Fig. 1).
This result supports the view that the pulsar high frequency radio emission 
originates from 
a region very close to the pulsar surface at a few percent of the 
light cylinder radius.

\begin{figure*}[t]
\epsfysize5cm
\rotate[r]{\epsffile[300 25 560 235]{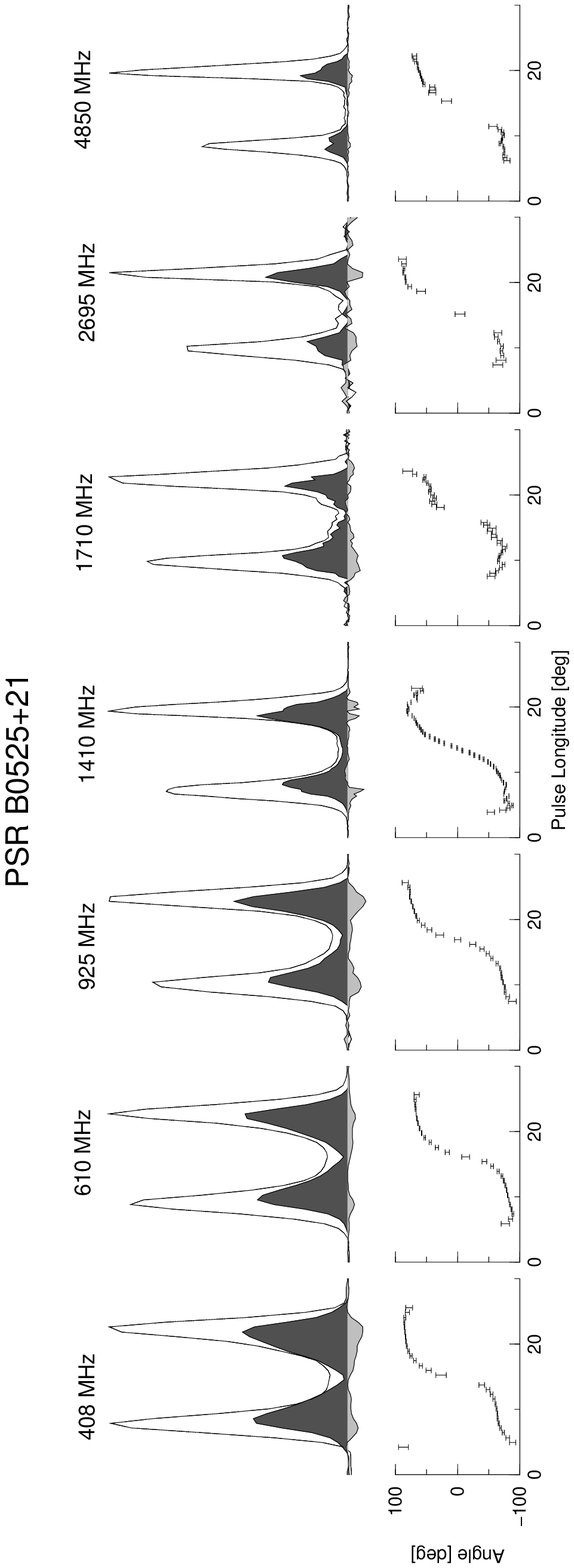}}
\caption[]{
Frequency development of the polarization profile of PSR B0525+21.
(profiles below 1 GHz are from Gould \& Lyne (1998), the others are 
Effelsberg data). 
For details see Fig. \ref{0355_freq}.
A nice double component profile with an PPA swing which
nearly perfectly agrees with the RVM. Significant contributions of 
circular polarization can be seen below the two intensity peak and correlate
well with deviations from the RVM. The narrow profile depolarizes well 
towards high frequencies.
}\label{0525plot}
\end{figure*}

\section{Discussion}\label{discussion}

We have analyzed our data together with high frequency data which
was already published by von Hoensbroech \& Xilouris (1997b) and 
compared the polarimetric
properties to those at lower frequencies. The low frequency data 
is from Gould \& Lyne (1998) and was accessed through the
European Pulsar Network (EPN) internet database 
\footnote{The EPN internet database can be accessed through: \it http://www.mpifr-bonn.mpg.de/pulsar/data/}.

\subsection{Typical polarimetric types of pulsars}\label{types}

The great variety of pulsars in pulse-shapes and polarization 
characteristics has always puzzled many researchers. Some authors proposed 
empirical  classification concepts
(e.g. Rankin 1983; Lyne \& Manchester 1988). 
These classification schemes are based on the
different properties of {\it core} and {\it conal} components: 
The core components are usually located close to the profile centre 
and tend to show a moderate degree of circular 
polarization and an unsystematic PPA swing. 
Cone components show hardly any circular 
and a moderate degree linear polarization. The PPA
swing is more systematic and can usually be modelled with the RVM. 
The spectral index 
($si$ in $S \propto \nu^{-si}$) for the core components
is higher than for the cone components (see also Kramer et al. 1994). 
Therefore, at high frequencies cone components become 
increasingly visible, if not dominant. Finally, core components 
tend to be more stable and features like mode changing -- temporal
changes of the average profile -- and
drifting subpulses -- stable subpulse features, which drift 
systematically with respect to the pulse phase through successive 
individual pulses, see e.g. Manchester et al. (1975) -- are rarely 
seen for these components.

The shape of a pulse profile then mainly depends on the {\it impact 
angle} and on the pulsar-specific {\it source function}, 
which represents the (patchy?) 
filling function of the emitting tube (e.g. Lyne \& Manchester 1988;
Manchester 1994).
If the impact angle is small enough, the line of sight possibly cuts through 
the central beam and the corresponding core  component becomes visible.

Our pulsar sample at 4.9 GHz generally fits well in these concepts. 
Nevertheless, we find a few groups of objects whose properties differ 
from what is expected from these classification systems. As these are
not just individual objects, but form groups with similar characteristics, 
it is suggested that additional (or different) effects can play a role
for the creation of intensity and polarization profiles of radio pulsars.
As some authors already pointed out, there is an evolutionary aspect to
keep in mind. So young pulsars tend to be highly polarized (e.g. \cite{Ml94})
and the cone--dominated ones are usually old (e.g. \cite{R83}).

\subsubsection{0355+54-Type}\label{0355}

\begin{figure*}[t]
\epsfysize5cm
\rotate[r]{\epsffile[300 25 560 235]{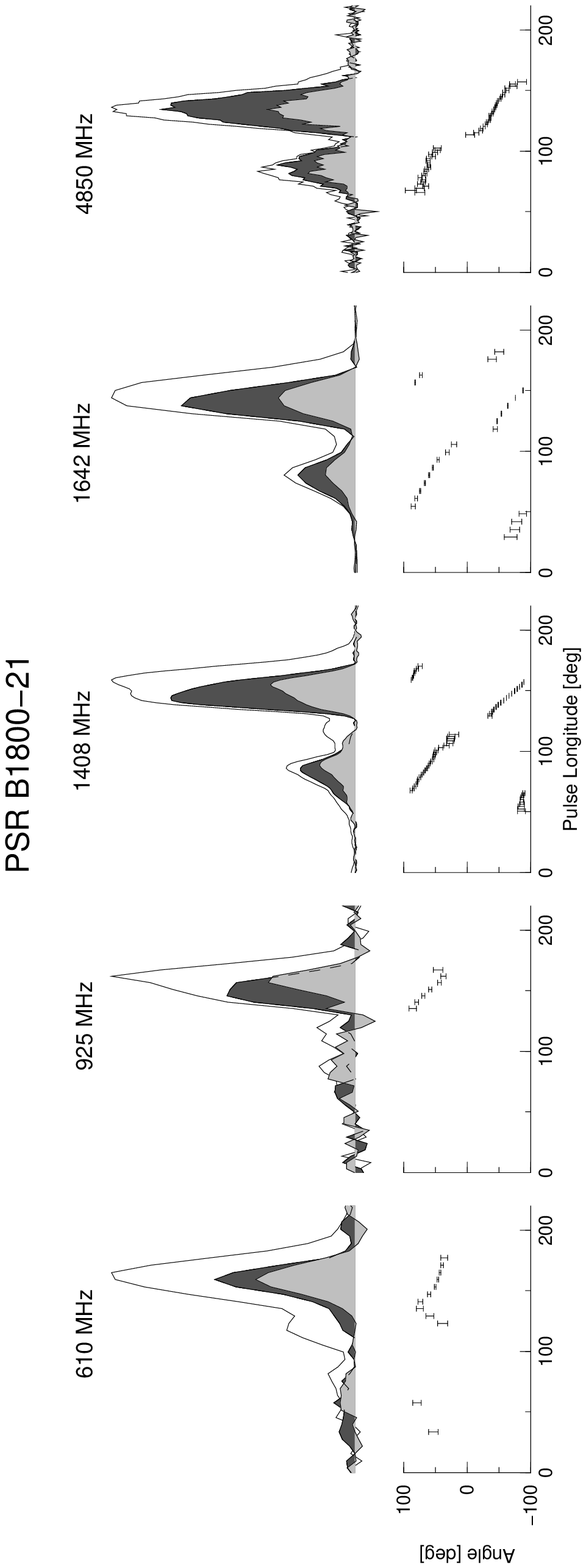}}
\caption[]{Frequency development of the polarization profile of 
PSR B1800$-$21.
(The 4850 MHz profile is Effelsberg data, the other ones are from 
Gould \& Lyne (1998)). 
For details see Fig. \ref{0355_freq}.
The wide profiles show strong linear and circular polarized contributions
over most parts of the pulse which increase towards high frequencies.}
\label{1800type}
\end{figure*}

This group of pulsars, which is represented by the pulsar B0355+54, 
consists of objects which share the following features: 
One component is highly linearly polarized ($\Pi_{\rm L}>60\%$) 
whereas the rest of the profile
shows only little or moderate polarization. The highly polarized
component has a flatter spectral behaviour than the profile
as a total, thus dominating it at high frequencies.
(see Fig. \ref{0355_freq}). 
Examples for such pulsars are PSRs B0355+54, B0450+55, B0599$-$05, B0626+24,
B1822$-$09, B1842+14, B1935+25 and B2224+65.
The existence of pulsars with a highly polarized leading component 
has already been noted in several papers (e.g. Manchester et al. 1980;
Morris et al. 1981a), 
but within this group we find also 2 pulsars where the polarized 
component is trailing.
Interestingly, these pulsars seem to have preferably two main components. 
They have usually been classified as {\it half cones} 
(e.g. Lyne \& Manchester 1988; Rankin et al. 1989), 
meaning that only one half of the
polarized cone can be seen. This view has been supported by the fact that
the PPA follows a shallow change during the polarized part of the 
profile (corresponding to an outer part of the emission cone) and 
a steeper change during the unpolarized component (corresponding to
an inner part). 
The surprising fact is that, if they were really half cones, one
would expect to find even a larger number of pulsars with two
highly polarized components, enclosing one unpolarized part. 
We are not aware of any such example, 
thus this observation puts doubt on the interpretation of these pulsars 
as half cones. 

\subsubsection{0525+21-Type}

These are the ``textbook''-pulsars.  Classified as conal doubles (\cite{R83})
they represent the traditional hollow cone model. The two components are 
moderately linearly polarized and the PPA follows nearly perfectly 
the RVM. Between the two components they usually show a core-type
bridge (see Fig. \ref{0525plot}). 
Examples for such pulsars are the PSRs B0148$-$06, B0301+19, B0525+21, 
B0751+32, B0917+63, B1039$-$19 and B1938$-$04.
But there are some remarkable facts to mention about these pulsars.
They often show a certain degree of circular polarization, which is 
usually associated with the conal components and not with the bridge (as
one would expect from the classification). 
They depolarize strongly 
towards high frequencies. As they are usually weak objects with
a steep spectral index, it is relatively difficult to
obtain high frequency polarization profiles. 

Interestingly these pulsars are all relatively old objects which
are located close to the ``death-line'' in the $P- \dot P$ diagram. 
The periods are
long and the loss of rotational energy $\dot E$ is low.
Only few objects of this kind are known. This suggests that the relative 
geometry of pulsar and observer is only partly responsible for the 
profile morphology. It seems that a major role -- both for the 
profile morphology and the polarization characteristics -- is played by
the age $\tau$ and therefore by intrinsic pulsar parameters, 
such as $\dot E$, which again determine
the ambient plasma conditions (see also Sect. \ref{corrl}).
The ability to emit polarized high frequency radiation in general,
and core emission in particular, is obviously reduced for such pulsars.
As these objects are usually also the oldest stars, their profile
morphology rather appears to be a (final) evolutionary stage than the 
product of a random geometry (see also \cite{R83}).

\begin{figure*}[t]
\epsfysize5cm
\rotate[r]{\epsffile[300 25 560 235]{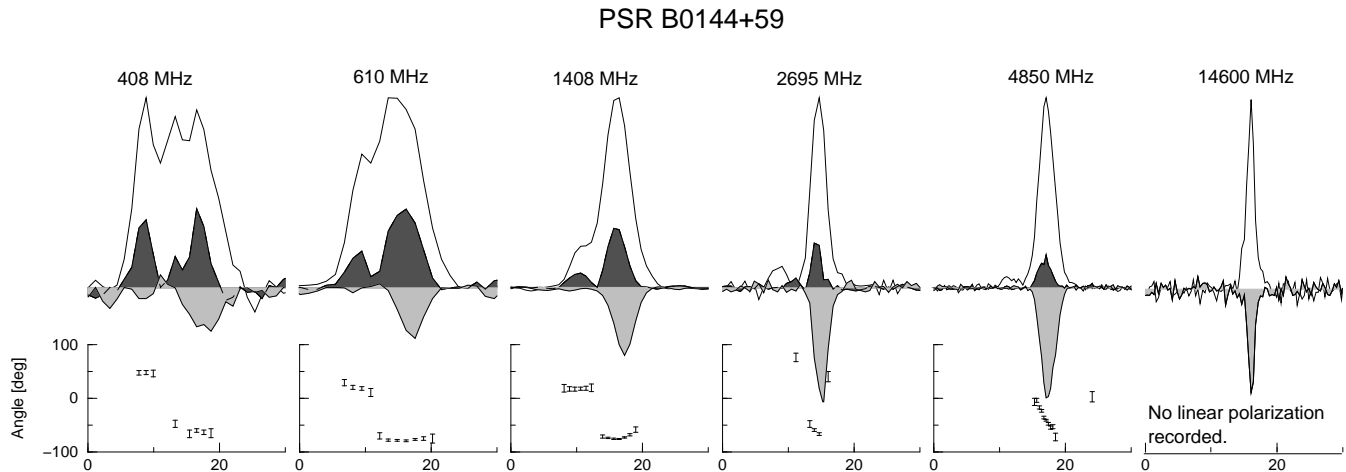}}
\caption[]{Frequency development of the polarization profile of PSR B0144+59. 
The profiles below 2 GHz are from Gould \& Lyne (1998), the other 
profiles are previously unpublished Effelsberg data. Note, that for the 
14.6 GHz profile, only circular intensities were measured. 
The degree of circular polarization
increases strongly towards high frequencies. The frequency development
of this pulsar is just opposite from what one would expect for a core single
($=S_t$) pulsar.}\label{0144_freq}
\end{figure*}

\subsubsection{1800$-$21-Type}\label{1800}

Another remarkable group of pulsars shares the following features:
They are young objects with a very high $\dot E$, a flat radio spectrum
behaviour, X-ray emission and radio-emission over a wide fraction of the
pulse period. They are nearly
fully polarized, linearly and circularly, and do not show any observable 
depolarization effects towards high frequencies (see Fig. \ref{1800type}). 
Examples are the pulsars B1800$-$21, B1823$-$13 and B1259$-$63 
(\cite{MJ95}). 

The behaviour of these pulsars is difficult to explain within the 
classification scheme, although parallels to
conal double pulsars have been drawn for B1800$-$21 by Wu et al. (1993)
and for B1259$-$63 by Manchester \& Johnston (1995). But there are major
differences. The 1800$-$21-type pulsars are
much stronger polarized and show so far no depolarization towards 
high frequencies, as it seems to occur strongly for the 
classical conal pulsars (\cite{R83}).
The high degree of polarization suggests that competing OPMs are
not active in most parts of the profile. So either only one polarization mode
is produced, or only one mode can propagate. Other differences are 
the general properties of these objects (see above). In every respect they
are opposite to the classical conal pulsars.

The fact that the polarization characteristics of this group are very unusual
and the pulsars share the above mentioned other unusual
properties, raises the question that there might be an intrinsic relation
between a high $\dot E$,  a shallow spectrum and a high polarization at 
high frequencies. We investigate this question in more depth in the
following section.

\subsubsection{0144+59-Type}\label{0144}

During our high frequency analysis we found five pulsars
which show an unusual polarization behaviour. Their degree of circular
polarization {\it increases} strongly with frequency
(see Fig. \ref{0144_freq}). This is in
sharp contrast to the ``textbook''-sentence: ``Pulsars depolarize
towards high frequencies''. As we regard this effect as critical
for the emission physics, we will consider it separately in 
Sect. \ref{circ}. Here we will just discuss these stars in the
light of the classification system.

Looking through published data and our own data, we found
five examples for this behaviour and a couple of suspects (the lack
of high frequency data makes it difficult to confirm the suspects as
members of this group). The clear examples are PSRs B0144+59, B0320+39, 
B1737$-$30, B1913+10 and B1914+13.
Due to their strong circular polarization these pulsars would
be usually classified as core-pulsars, but especially the 
frequency development of the B0144+59-profile puts doubt on this
interpretation. At low frequencies, three
components can clearly be identified. In opposite behaviour to the 
prediction of the classification model, the moderately linearly
polarized outer components (outriders) weaken towards higher
frequencies and the strongly circularly polarized central 
component dominates the profile. In fact, if the frequency development
would be reversed, it would be a perfect example for a core-single ($S_t$) 
pulsar.

\begin{figure}
\epsfysize5cm
\rotate[r]{\epsffile[120 90 520 440]{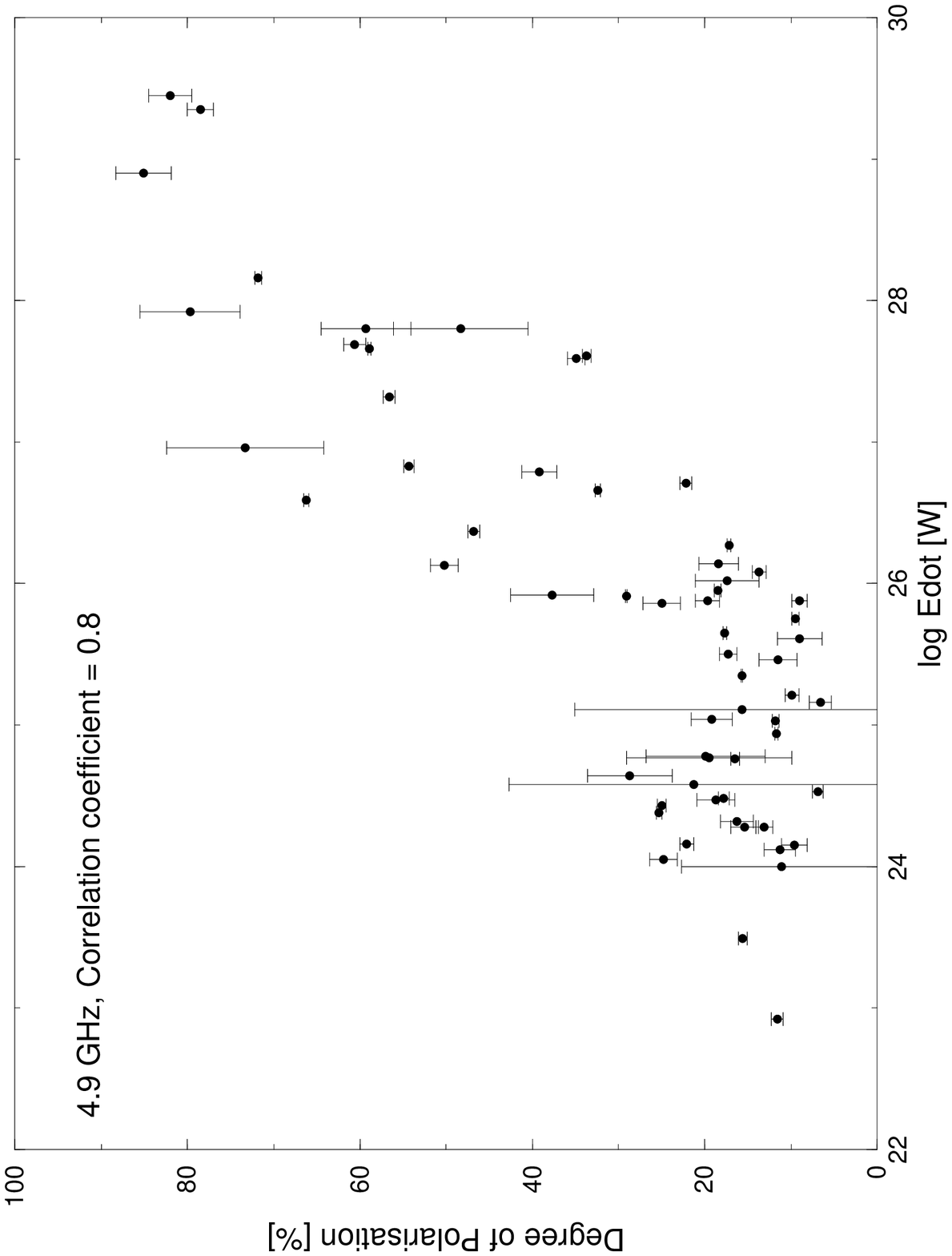}}
\epsfysize5cm
\rotate[r]{\epsffile[65 90 465 440]{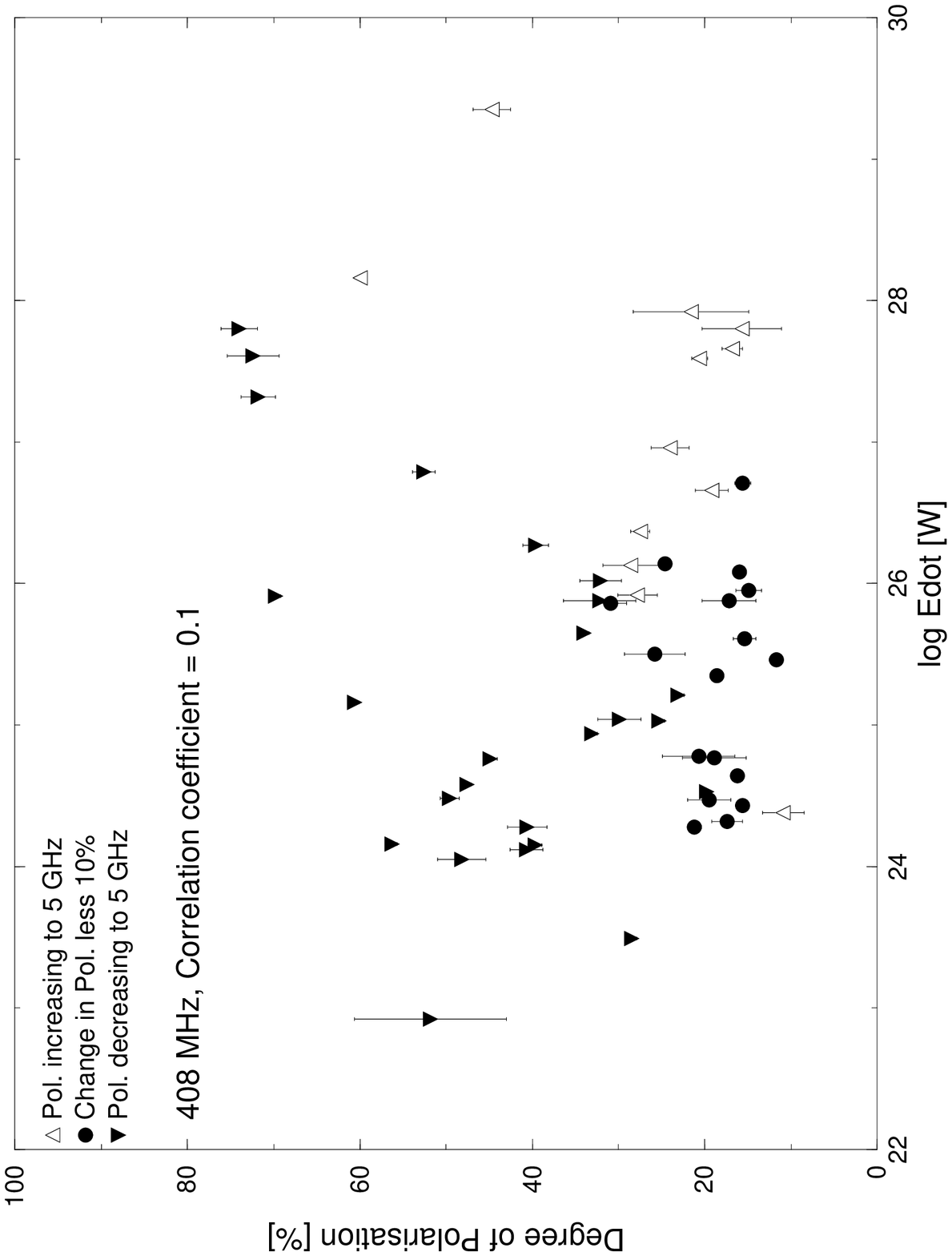}}
\vspace*{1cm}
\caption[]{Plot of the integrated degree of polarization versus the 
rotational energy loss $\dot E$. {\bf Upper plot:} Correlation
for 62 pulsars at 4.9 GHz This correlation is highly significant (see text). 
{\bf Lower plot} Same set of pulsars
at 408 MHz (5 pulsars are missing as no low frequency data was
available, 2 are recorded at 610 MHz). Obviously no statistically significant 
correlation is apparent. 
The degree of polarization {\it increases} for high $\dot E$
pulsars and {\it decreases} for low $\dot E$ pulsars towards high frequencies
(see text).
The low frequency data were taken from Gould \& Lyne (1998).}
\label{PEdot}
\end{figure}

\subsection{Correlations between polarization and other pulsar parameters}\label{corrl}

An important question within emission physics is, whether the polarization
properties 
correlate with any other pulsar parameters. Some researchers have already
noticed that the degree of polarization weakly correlates with $\dot P$
and anti-correlates with the period $P$ (Morris et al. 1981a; Gould 1994). 
But this was done at frequencies $<$ 3 GHz and not very significant.
The high frequency range on the other hand is of special interest for this 
question since the depolarization occurs in this regime. A possible
correlation could tell us something about the relevant parameters for the
depolarization process (which might actually be essentially the same
as the polarizing mechanism). To our knowledge two attempts have already 
been made to correlate the {\it depolarization index} (which is the
power-law-index of $\Pi_{\rm L}\propto \nu^\alpha$) to other pulsar 
parameters. Using 2.7 GHz data, Morris et al. (1981a) presented 
indications, that long period pulsars depolarize faster than short period 
ones. 
Xilouris et al. (1995) presented an anti-correlation 
between the depolarization index and the surface accelerating potential
$\Phi_\parallel=B_{12}/P^2$ ($B_{12}$ being the surfave magnetic field 
strength) for 29 pulsars, using 10.6 GHz data.
Also during the analysis of our data, it became clear, that highly 
polarized emission at 4.9 GHz can only be found with
short period pulsars. This corresponds to the relationship found
by Morris et al. (1981a) at 2.7 GHz. 

The degree of polarization therefore seems to correlate more or
less significantly with various pulsar parameters. We want to know, 
with which property the best correlation can be found,
applying the following procedure:
The two parameters which we can measure with the greatest accuracy
are the period $P$ and the period derivative $\dot P$. From these 
parameters we can derive other
pulsar properties, such as {\it characteristic age $\tau$, surface
magnetic field $B$, rotational energy loss $\dot E$ and the 
polar gap accelerating potential $\Phi_\parallel$}. 
The functional dependency of all these parameters is $\propto P^a
\cdot \dot P^b$, with different exponents $a$ and $b$.
In order to find a possible correlation 
with one of the pulsar properties, we correlated the degree of polarization  
with the above function and varied $a$ and $b$. 
The quality of the correlation was then tested using the Pearson-test and
the Spearman-rank test (e.g. \cite{NumRes}) independently. 

Interestingly we found the best correlation not only for one, but for 
a number of combinations of $a$ and $b$.
The significance of the correlation is 
$5.5 \sigma$, corresponding to a 99.8\% probability that the observed 
correlation is true. 
All combinations of $a$ and $b$ where this best correlation was found,
 fulfil the condition: $a/b =-3$.
There are two pulsar properties, which match this condition:
The  rotational energy loss $\dot E \propto \dot P/ P^3$ and the
polar gap accelerating potential
$\Phi_\parallel \propto (\dot P / P^3)^{1/2}$ 
(see Fig. \ref{PEdot}, upper plot).

It is therefore suggested, that these two (intrinsically closely related)
parameters strongly influence the polarization properties of
the pulsar radio emission. 

A similar correlation could not be detected for nearly the same set of
pulsars observed at low frequencies (see Fig. \ref{PEdot}, lower plot). The 
integrated degree of polarization seems to be randomly distributed
with respect to $\dot E$. Comparing to high frequencies, it becomes
obvious that mainly highly polarized low-$\dot E$ pulsars show significant
depolarization, whereas weakly polarized high-$\dot E$
pulsars show an increasing degree of polarization to high frequencies.
This indicates that pulsars with an increasing degree of polarization
are not just ``outriders'', but rather form a systematic group
within the pulsar population.

In the following we want to speculate in a straightforward fashion about
some possible physical implications of this correlation.
Under the assumption that the polarization is not strongly influenced
by any outer gap plasma effects, it
shows one important fact: When the plasma radiates the
radio emission, it somehow ``knows'' about the accelerating potential 
inside the polar gap respectively about the loss of rotational energy 
$\dot E$. 
Until the emission of the observed radiation no processes are allowed, 
which would result in the loss of this information.

As a possible influence of $\dot E$ on the radiating plasma, we regard the
following: A high $\dot E$ implies a 
high $\Phi_\parallel$. This again leads to a higher total kinetic
energy of the out-flowing plasma. This energy is given by 
\begin{equation}
E_{\rm kin}=\bar\gamma\cdot N\cdot m_e c^2 ~~~{\rm with}~~~ \bar\gamma = \frac{\int n(\gamma)\cdot\gamma\cdot d\gamma}{\int n(\gamma)
\cdot d\gamma}~.
\end{equation}
$n(\gamma)$ is the energy distribution function of the radiating plasma.
In order to increase the total energy, either the number of plasma-particles 
$N=\int n(\gamma) d\gamma$ or $\bar\gamma$ has to increase.
To obtain a larger $\bar\gamma$, the distribution function 
$n(\gamma)$ has to be changed either by shifting it to
higher energies or by ``flattening'' the function, thus getting a higher 
particle population at high $\gamma$'s.
It is unlikely that the whole function can be shifted to much higher 
energies, as pair-creation and emission of X- and $\gamma$-rays will always 
limit the $\gamma$-factor below certain values. In this context it is 
interesting to note that the X-ray luminosity correlates strikingly well 
with $\dot E$ (\cite{BT97}), which supports the above statement.

We could therefore conclude that either a higher plasma density or a 
flatter distribution function $n(\gamma)$ implies a higher
degree of polarization. There can be two reasons for this:
\begin{itemize} 
\item If the polarization is an intrinsic property of the
emission mechanism it could be concluded that the coherence increases
with the plasma density and therefore less independently radiating bunches
are superposed and depolarize the radiation. It is also possible that the 
polarization of the emission mechanism depends on $n(\gamma)$. In this case
the degree of polarization would depend on the shape
of $n(\gamma)$ insofar as a flatter distribution function results into
a higher degree of polarization.
\item The polarization may be caused or at 
least strongly influenced by a propagation effect. The influence would be
a different frequency dependent refractive index for the two natural
(and orthogonal) propagation modes. The frequency walk of these
refractive indices would then depend strongly on the plasma density or
$n(\gamma)$.
\end{itemize}

It might be interesting in this context to point to the fact that 
there is a slight correlation between the spectral index and $\dot E$, 
such that pulsars with a high $\dot E$ often have a flatter spectrum. 
For example Lorimer et al. (1995) presented 
an anti-correlation between the spectral index and the characteristic age, 
which again is anti-correlated with $\dot E$.
Especially those pulsars which have a high $\dot E$ {\it and} 
a high polarization 
(1800$-$21-type, Sect. \ref{1800}) have a very flat spectrum. 
This corresponds to the behaviour of the highly polarized profile
components of the 0355+54-type pulsars (see Sect. \ref{0355}), which also 
have a significantly flatter spectrum than the rest of the profile
(see Fig. \ref{0355_freq}).
The spectral behaviour of non-thermal radiation mechanisms mainly 
depends on the 
energy distribution function of the emitting particles. If this
function is flat, thus more particles have higher energies, 
then the resulting spectrum is flat as well.
Therefore it seems that especially the {\it shape} of the 
energy--distribution function $n(\gamma)$ is influenced by $\dot E$.

Concluding one should note that these observations suggest an intrinsic
relationship between a high $\dot E$, a high degree of polarization, 
and a flat spectrum. This is independent
of any core or conal classification of these stars respectively their profile
components. It rather looks as if pulsars with a moderate or high $\dot E$
(and therefore also high $\Phi_\parallel$) can develop certain
regions on their polar caps, where they direct most of their available 
energy into the plasma by flattening the distribution function $n(\gamma)$. 
This plasma then
emits a highly polarized radiation with a comparably flatter spectrum than
the rest of the radiating plasma. One could even speculate that, 
if such a region develops, it reduces the plasma energy above other
parts of the neutron star, leading to an anti-correlation in the 
occurrence of certain
components in single pulse observations, as it is observed e.g. 
for PSR B1822$-$09 (\cite{G94}).

\subsection{Frequency development of circular polarization}\label{circ}

One fundamental piece in the theory of pulsar radiation -- where no 
consensus has been found yet -- is the question whether the polarization
arises from the emission process itself or whether it is produced by
a birefringinal propagation effect (e.g. Melrose \& Stoneham 1977; 
Cheng \& Ruderman 1979; Barnard \& Arons 1986; McKinnon 1997). 
The latter case would allow a larger number of considerable 
emission mechanisms as also processes which are intrinsically 
{\it unpolarized} could be considered.

Qualitatively this would correspond to a separation of the original 
radiation into two orthogonal natural modes which both experience different
frequency dependent indices of refraction. 
A spatial or temporal separation of these two modes could then
result into OPMs, a differential absorption into the domination
of one mode above the other. 
The properties of the two modes would be frequency dependent. If the plane
of polarization develops an angle to the magnetic field lines during the
propagation, the ratio of the two great axis of the polarization ellipse could
undergo changes which would result into a transition of linear polarization
into circular. 
Some authors have therefore already suggested, that if propagation effects
play a role in the pulsar magnetosphere, they should result into an 
increasing degree of circular polarization with frequency 
(Melrose \& Stoneham 1977; Cheng \& Ruderman 1979). 
The non-existence of corresponding 
observations was attributed to a possible competing depolarization 
mechanism. However authors as Manchester et al. (1980) noted already 
that some pulsars like Vela show a slightly
increasing circular contribution. 

The frequency-development of the polarization profile of PSR B0144+59
is to our knowledge the first significant observation of this
effect (see Fig. \ref{0144_freq} and Sect. \ref{0144}). The degree of circular polarization 
increases constantly with frequency to a very high value 
($\Pi_{\rm C} \ga 50 \%$),
whereas the total degree of polarization 
($\Pi_{\rm tot} = \sqrt{\Pi_{\rm L}^2 + \Pi_{\rm C}^2}$) is 
approximately constant 
above $\sim 1$ GHz. It seems unlikely that an intrinsic mechanism 
other than a propagation effect could result in such a frequency dependence.
We therefore suggest that the magnetosphere can have a birefringing 
quality, which transforms linear polarization into circular, similar 
to a $\lambda /4$-plate.

Another puzzling aspect about the frequency development of this pulsar 
is the change of the PPA. One of the basics in the knowledge of 
pulsar polarization is the stability of the polarization curve -- which
eventually lead to the conclusion that the shape of this curve is determined
by the geometry of the pulsar. The only known qualitative changes are 
connected to the (frequency dependent) occurrence of OPMs. But in 
Fig. \ref{0144_freq} we observe a smooth steepening of the PPA curve.
The same effect is also observed for PSR B1737$-$30 another pulsar 
of this class. The classical interpretation for this change would be 
that the line of sight changes with frequency. An alternative interpretation 
could be that the same propagation effect which transforms linear into 
circular polarization, leads to a differential rotation of the 
plane of polarization.

Unfortunately only little data about such pulsars is available, so no
general statement can be drawn yet. But additional observations will be 
carried out soon.

\section{Summary and Conclusions}

We have analysed average radio pulsar polarization
profiles at high frequencies and present 32 previously unpublished pulsar 
polarization profiles measured at a frequency of 4.9 GHz. 
The profiles are also available in EPN-format (\cite{L98}) through
the EPN internet database (see Sect. \ref{discussion}).
The properties of the whole available
set of 4.9 GHz average polarization profiles in general and those of 
individual pulsars in special were compared to lower frequencies. 

Investigating the average polarization profiles of individual 
pulsars with particular respect to their classification in the scheme
of Rankin (1983), we found groups of pulsars which deserve additional
attention. 
\begin{itemize}

\item Some pulsars, such as PSR B0355+54, have components with
very different polarimetric and spectral properties within the
same profile. One component
is nearly fully polarized and has a flatter spectrum than the
profile as a whole, thus 
dominates the profile at high frequencies. The rest of the 
profile is hardly polarized and dominates only at low frequencies.
As all eight pulsars which form this group so far, show a similar
frequency-dependence, it is suggested that an intrinsic correlation
exists between a high degree of polarization and a flat spectrum.
All these pulsars have been classified as half-cones. Although this
classification is tempting, it is important to note that not a single
full-cone with similar properties could be found. This indicates that
a more general process takes place than just an occasional
lack of flux at the position where the line of sight cuts the 
cone for the second time.

\item There are three young pulsars (B1800-21, B1823-13 and B1259-63) 
with a very high loss of rotational
energy $\dot E$ and a very flat spectrum. These pulsars are nearly fully
polarized and do not show any significant depolarization.
This behaviour shows similarities
to the above mentioned highly polarized components of the 0355-like
pulsars. Both groups show a correlation between high polarization
and a flat spectrum.

\item We found a number of objects which show a circular polarization, which
strongly {\it increases} with frequency. This is in sharp contrast to the
known frequency-dependence of pulsar polarization.
As the linear polarization decreases simultaneously, it is suggested 
that a propagation-effect similar to a $\lambda/4$-plate is active.
If confirmed, this could indicate, that propagation effects 
influence the polarization within the magnetosphere. 
Additionally, these pulsars fit hardly into the classification scheme.
PSR B0144+59 for instance shows precisely the opposite frequency-development
to a $S_t$-pulsar (see Fig. \ref{0144_freq}).

\end{itemize}

We would like to point out again the possible role
of pulsar evolution on the polarization profile. Within the empirical
model for pulsar emission, the profile shape and the polarization
properties is nearly exclusively determined by the viewing geometry 
of pulsar and line of sight and the activity in the different parts of the
magnetosphere. But, as it was already noted by 
Rankin (1983), the classical conal double pulsars (the ``textbook-pulsars'',
0525+21-type)
are without exception very old stars which lie close to the
``death-line'' in the $P-\dot P$-diagram. Contrarily the 1800$-$21-like
pulsars mentioned above are very young pulsars with very different
properties. 
For future work it appears to be important to focus stronger on the role 
of evolution for polarization profile shapes.

Analysing the general properties of pulsar polarization profiles at this
frequency of 4.9 GHz, we found a significant correlation between the total 
degree of polarization with $\dot E$ and the $\Phi_\parallel$
at the polar gap respectively  (see Fig. \ref{PEdot}, upper plot). 
Such a correlation does not exist at
lower frequencies. The pulsars at low frequencies  rather form groups 
which have a decreasing (highly polarized, low $\dot E$ pulsars) and an 
increasing (weakly polarized, high $\dot E$ pulsars) degree of 
polarization to high frequencies. 
Observations at high radio frequencies therefore yield additional 
information on the emission physics which is not seen at lower frequencies.
This correlation confirms the relation between the depolarization index
and $\Phi_\parallel$ at 10.5 GHz, which was presented by Xilouris et al. (1995).

The large differences in the degree of polarization between 
individual pulsars indicate that the ambient physical conditions in the 
respective emitting region differ significantly among them. 
As we can see from the correlation, this depends on $\dot E$. 
$\dot E$ again is closely correlated
to the polar gap $\Phi_\parallel$. As a high degree of polarization seems 
to be correlated to a flatter spectrum (see above), we speculate that a 
high $\Phi_\parallel$ could induce a flatter energy distribution
function of the radiating plasma. 

\onecolumn
\begin{figure}[C]
\epsfysize23cm
\epsffile[25 70 525 770]{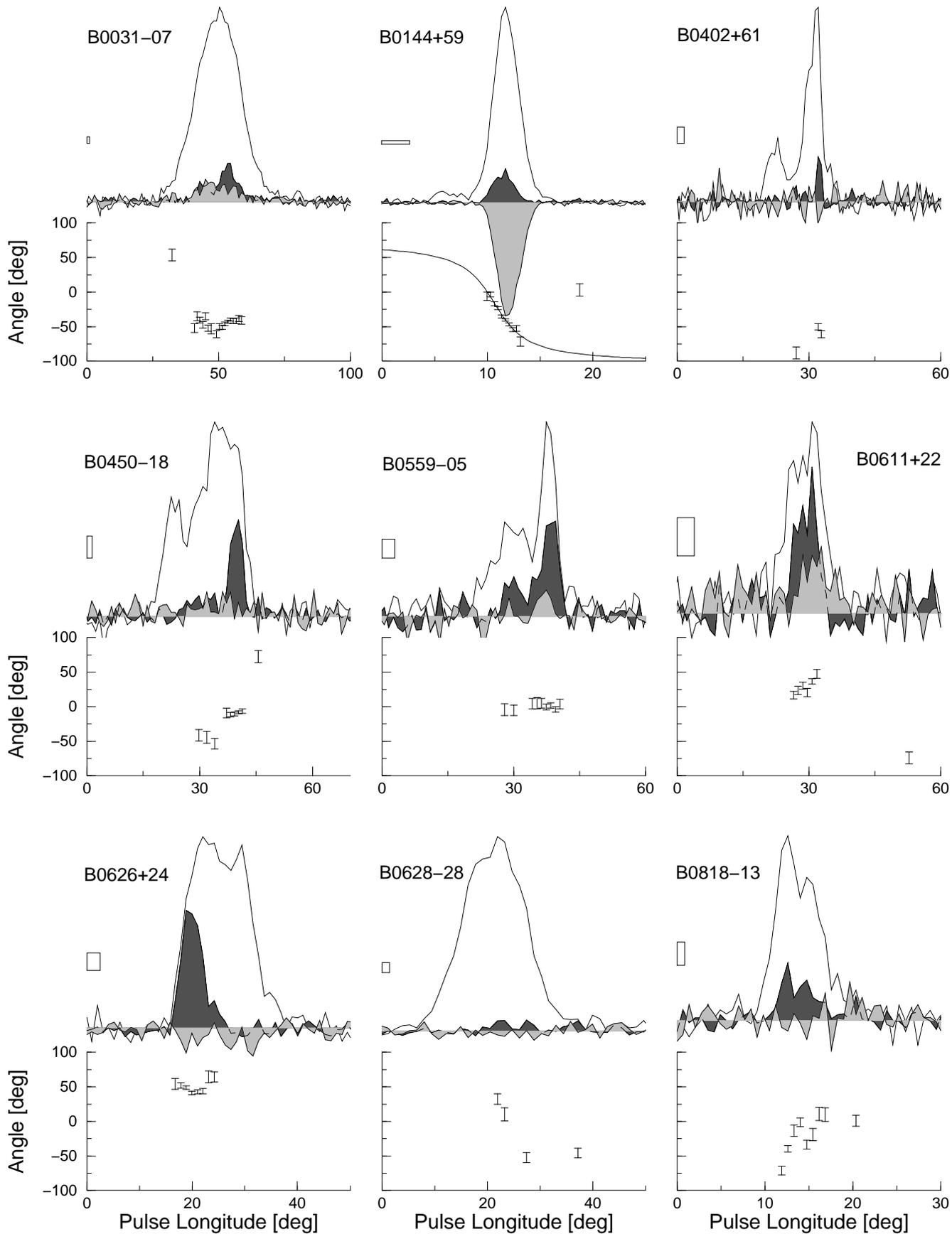}
\caption{Pulsar polarization profiles at 4.85 GHz.
The dark-shaded 
area represents the linear, the light-shaded area corresponds to the circularly
polarized intensity ({\it positive} $\hat =$ left-hand, 
{\it negative} $\hat =$ right-hand).
 Total power is represented by the unshaded solid line. The error-box has a
height of 2 $\sigma$ and a width corresponding to the effective 
time-resolution (see caption of Table~1).
When it was possible, the RVM was fitted to the angle (e.g. for B0144+59).}
\label{data1}
\end{figure}
\twocolumn

\onecolumn
\begin{figure}[C]
\epsfysize23cm
\epsffile[25 70 525 770]{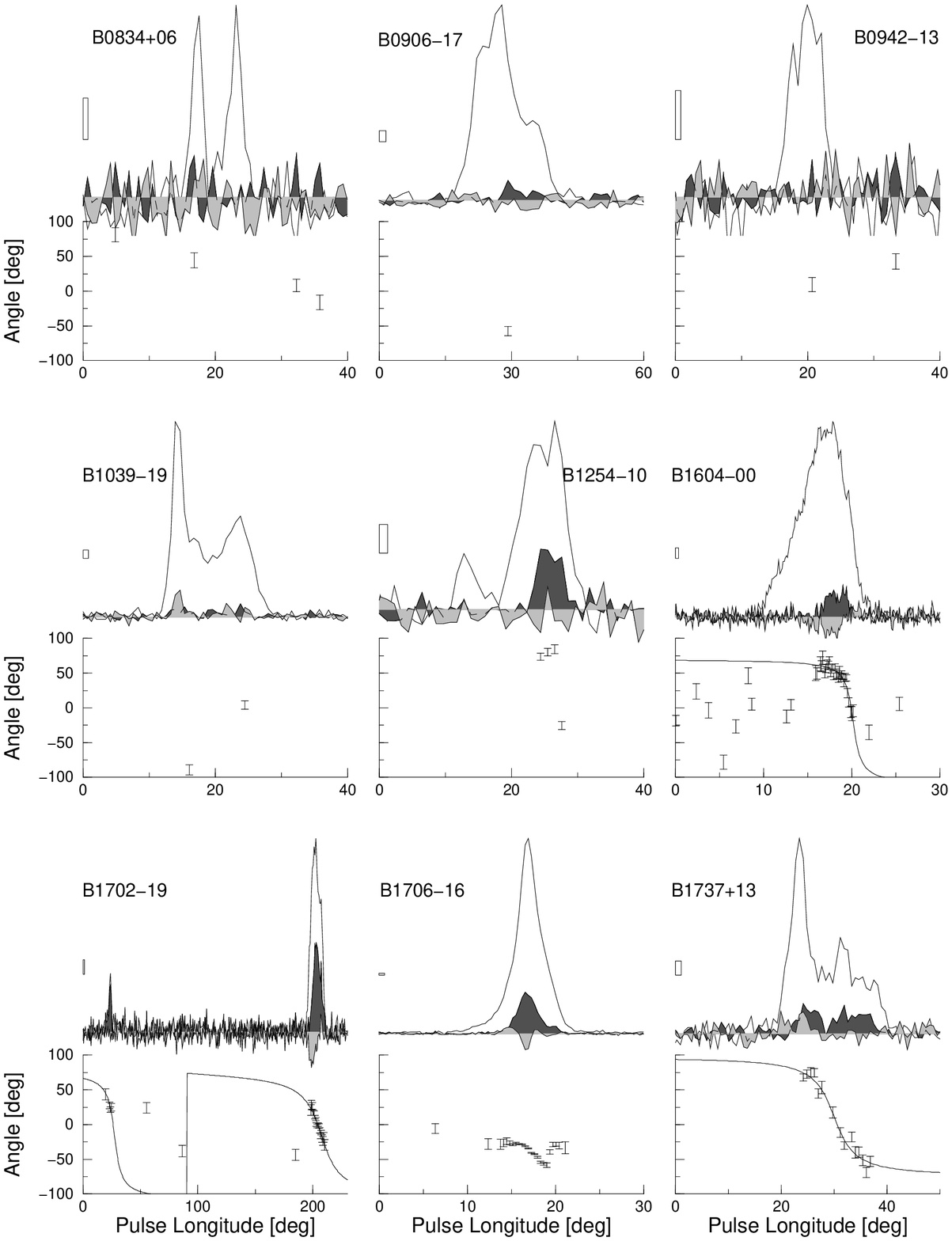}
\caption{Pulsar polarization profiles at 4.85 GHz. For details
see caption of Fig. 8.}
\label{data2}
\end{figure}
\twocolumn

\onecolumn
\begin{figure}[C]
\epsfysize23cm
\epsffile[25 70 525 770]{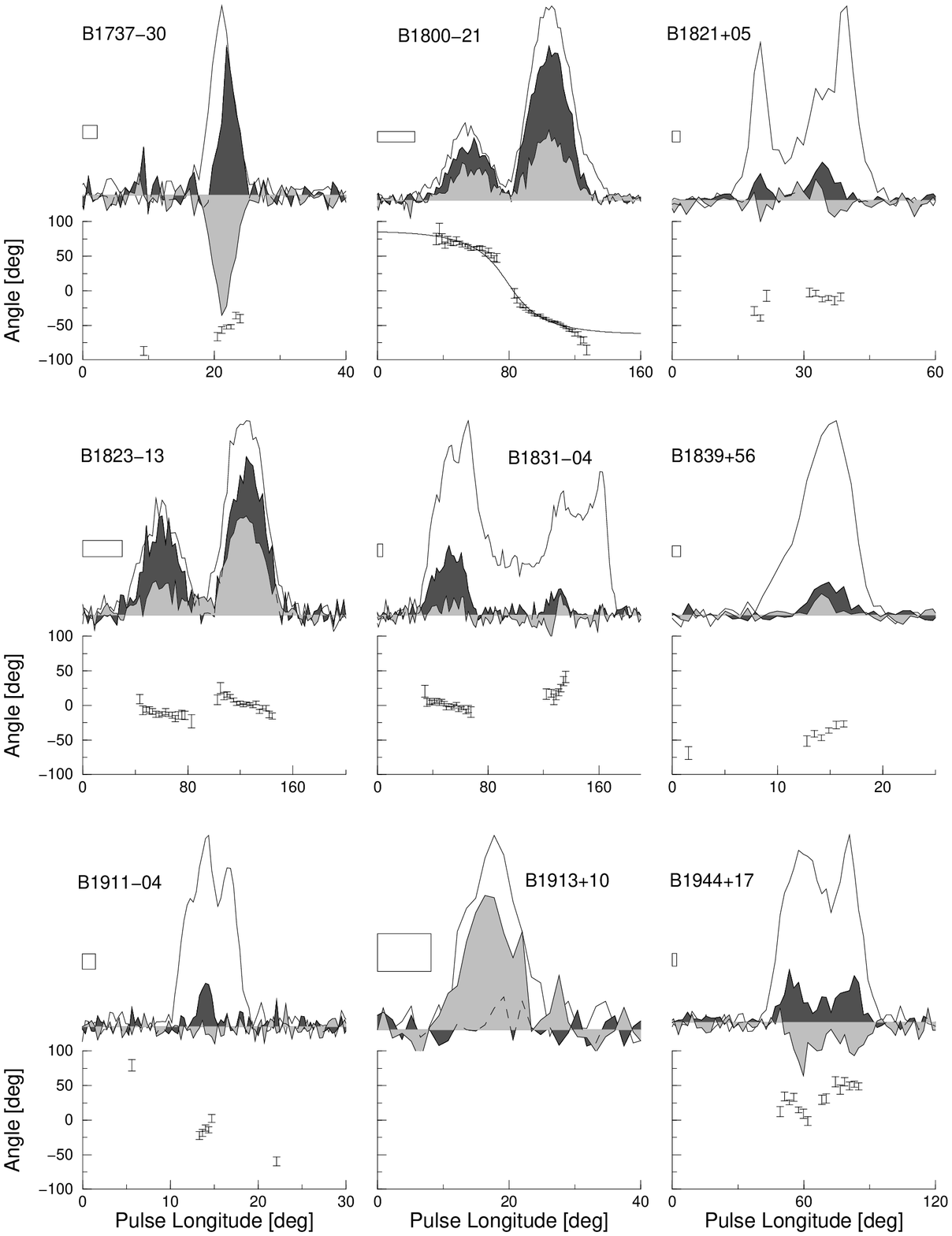}
\caption{Pulsar polarization profiles at 4.85 GHz. For details 
see caption of Fig. 8.}
\label{data3}
\end{figure}
\twocolumn

\onecolumn
\begin{figure}[T]
\epsfysize11.5cm
\epsffile[25 220 525 570]{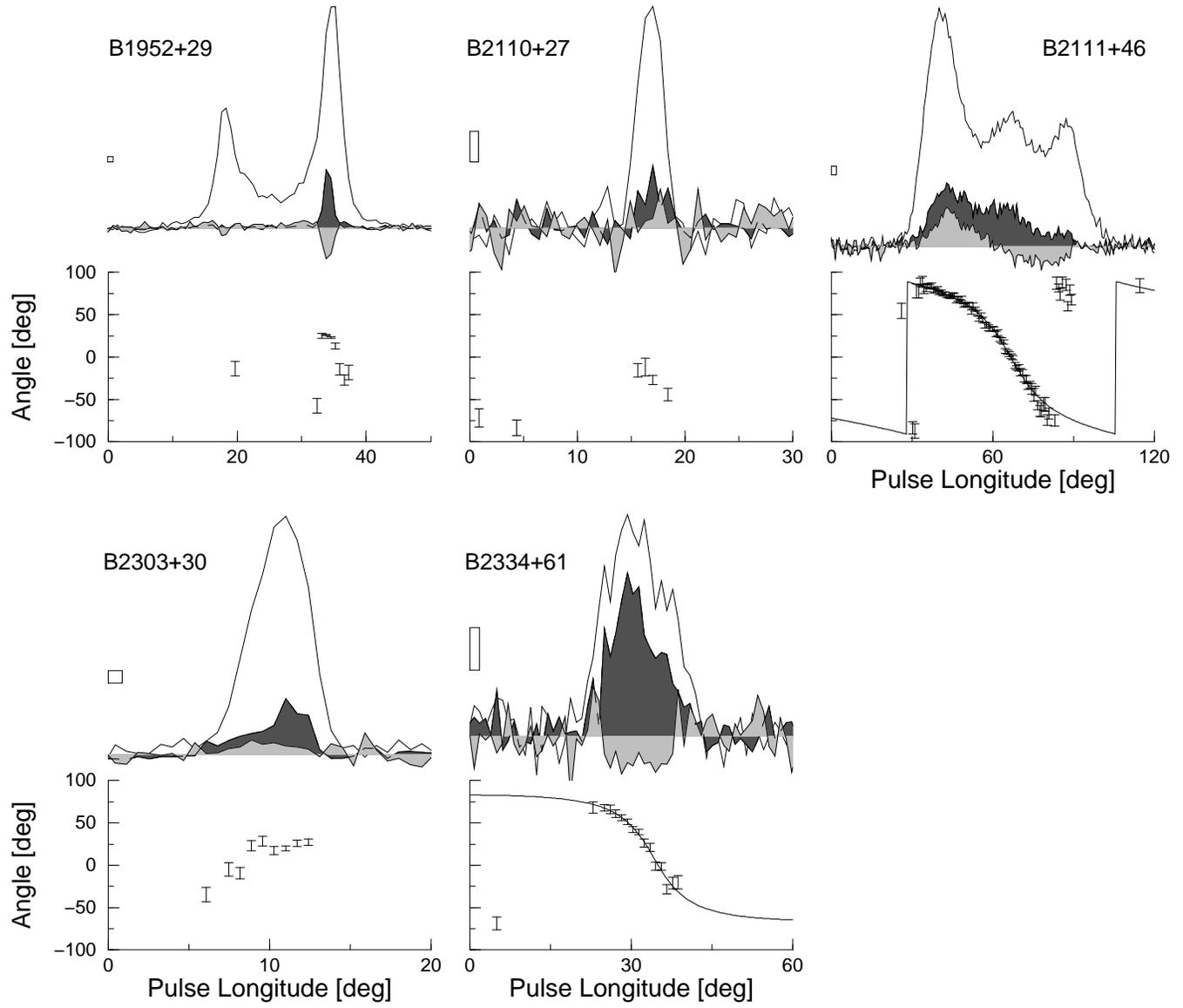}
\vspace*{-2cm}
\caption{Pulsar polarization profiles at 4.85 GHz. For details see caption of Fig. 8.}
\label{data4}
\end{figure}
\twocolumn

\acknowledgements  We would like to thank R. Wielebinski, R.T. Gangadhara, 
A. Jessner, M. Kramer, H. Lesch, D. Lorimer, W. Sieber and K.M. Xilouris 
for support and 
helpful comments on this paper. For their help with observation
we thank O. Doroshenko and C. Lange.
We also thank A.G. Lyne and M. Gould for providing us with
lower frequency polarimetry data through the EPN internet database
(supported by the European Commission under the HCM Network Contract 
Nr. ERB CHRX CT960633, i.e. the {\it European Pulsar Network}).
This paper was partially supported by the Polish
State Committee for Scientific Research Grant 2 P03D 015 12.

\end{document}